\documentclass[
 reprint,
 amsmath,
 amssymb,
 aps,
 prapplied,
 superscriptaddress,
 floatfix,
 longbibliography
]{revtex4-2}

\usepackage{graphicx}
\usepackage{xcolor}
\usepackage{hyperref}

\graphicspath{{figures/}}

\newcommand{\dd}{\mathrm{d}}

\hypersetup{
  colorlinks=true,
  linkcolor=blue,
  citecolor=blue,
  urlcolor=blue,
  pdftitle={Quantitative infrared absorption imaging of vapor concentration fields during evaporation},
  pdfauthor={Zihao Song, Weitian Wang, Chao Sun, and Xing Chao}
}

\begin{document}

\title{Quantitative infrared absorption imaging of vapor concentration fields during evaporation}

\author{Zihao Song}
\affiliation{New Cornerstone Science Laboratory,
Center for Combustion Energy,
Key Laboratory for Thermal Science and Power Engineering of Ministry of Education,
Department of Energy and Power Engineering,
Tsinghua University, Beijing 100084, China}

\author{Weitian Wang}
\affiliation{New Cornerstone Science Laboratory,
Center for Combustion Energy,
Key Laboratory for Thermal Science and Power Engineering of Ministry of Education,
Department of Energy and Power Engineering,
Tsinghua University, Beijing 100084, China}
\affiliation{Department of Mechanical and Automation Engineering,
The Chinese University of Hong Kong,
Hong Kong SAR, China}

\author{Chao Sun}
\email{chaosun@tsinghua.edu.cn}
\affiliation{New Cornerstone Science Laboratory,
Center for Combustion Energy,
Key Laboratory for Thermal Science and Power Engineering of Ministry of Education,
Department of Energy and Power Engineering,
Tsinghua University, Beijing 100084, China}
\affiliation{Department of Engineering Mechanics,
School of Aerospace Engineering,
Tsinghua University, Beijing 100084, China}

\author{Xing Chao}
\email{chaox6@tsinghua.edu.cn}
\affiliation{New Cornerstone Science Laboratory,
Center for Combustion Energy,
Key Laboratory for Thermal Science and Power Engineering of Ministry of Education,
Department of Energy and Power Engineering,
Tsinghua University, Beijing 100084, China}

\date{\today}

\begin{abstract}

Quantitative full-field measurements of vapor concentration remain challenging in evaporating flows, particularly within the gas layer adjacent to a receding liquid--gas interface.
Here, we develop and quantitatively assess an infrared absorption imaging method for spatially resolved measurements of vapor mole-fraction fields during evaporation.
Broadband mid-infrared attenuation is converted into vapor mole fraction using a spectrally weighted absorption coefficient, and a vertically oriented Hele--Shaw cell is employed to establish a controlled, nearly one-dimensional evaporation benchmark. The imaging system provides a calibrated object-plane sampling of $30.5~\mu\mathrm{m\,pixel^{-1}}$. The measured two-dimensional fields first confirm that transverse concentration variations are small within the central measurement region, supporting reduction of the data to axial concentration profiles. A canonical Stefan evaporation problem is then used as a stringent field-level benchmark.
Five profiles spanning different retained gas-phase interval lengths and three hydrocarbon liquids are compared with theoretical predictions using only the measured interval length and endpoint concentrations; no interior concentration data are fitted. At the higher vapor loading of \textit{n}-pentane, the measurements resolve the nonlinear concentration profile predicted by Stefan transport and distinguish it from the linear diffusion limit. At the lower vapor loadings of \textit{n}-hexane and \textit{n}-heptane, the profiles progressively approach that dilute limit. The measured profiles agree closely with the Stefan predictions, with normalized root-mean-square deviations of $0.0055$--$0.0185$, and collapse onto the predicted identity relation after the logarithmic Stefan transformation. These results demonstrate that infrared absorption imaging can quantitatively recover both the magnitude and spatial structure of vapor concentration fields over the tested conditions. The method therefore provides a full-field and tracer-free diagnostic for investigating evaporating flows in geometries and regimes for which analytical concentration fields are unavailable.

\end{abstract}


\maketitle

\section{Introduction}
\label{sec:introduction}

Evaporation is accompanied by coupled transport of mass, momentum, and energy in both the liquid and gas phases, and plays an important role in a wide range of natural and technological processes, including droplet drying~\citep{deegan1997capillary}, coating~\citep{Routh2013}, thermal management~\citep{Plawsky2014,Wakata2023Leidenfrost}, combustion~\citep{Law1982}, and additive manufacturing~\citep{lohse2022fundamental}.
Despite the apparent simplicity of evaporation from a quiescent liquid surface, quantitative characterization of the vapor phase remains experimentally challenging.
In particular, the vapor concentration immediately above the interface determines the local mass-transfer rate and mediates the coupling between interfacial phase change and gas-phase transport.
Spatially resolved measurements of this concentration field are therefore valuable for testing transport models and for studying more complex evaporating flows.

Most experimental studies of evaporation infer the mass-transfer rate from liquid-side quantities, such as the temporal variation of droplet volume, interface position, or total sample mass~\citep{Zeng2023,Wakata2024}.
These measurements provide an integral evaporation rate but contain limited information about the structure of the gas-phase concentration boundary layer.
Non-intrusive optical diagnostics have enabled spatially resolved measurements in gas flows that are difficult to access using mechanical probes, including measurements of velocity fields~\citep{Adrian1991,Danehy2022FLEET,Song2026Microjets} and scalar quantities~\citep{Borggren2017TLAF,Song2026Tin}. For quantitative measurements of vapor concentration specifically, several optical approaches have been applied.
Schlieren and interferometric techniques can visualize refractive-index variations associated with vapor concentration gradients~\citep{Dehaeck2014}, however, the refractive-index signal is not intrinsically species-selective, and quantitative interpretation may be further complicated by simultaneous temperature variations and by the indirect relation between the measured optical signal and species concentration.
Laser-induced fluorescence and related tracer-based methods provide improved spatial resolution, but require suitable fluorescent species or tracers and may be affected by quenching, photophysics, and calibration uncertainties~\citep{Schulz2005,Ehn2017}.
Laser absorption spectroscopy, in contrast, provides an intrinsically species-selective and quantitative relation between optical attenuation and molecular concentration and has been widely used for gas-phase species measurements~\citep{Goldenstein2017IR,Wang2019LAS,Zhu2022Comb,Wang2025HPSDS}.
However, conventional absorption measurements are commonly implemented along individual lines of sight or at discrete measurement locations, rather than as instantaneous two-dimensional concentration fields. 
Recent studies have extended laser absorption spectroscopy to spatially resolved measurements through beam scanning and tomographic reconstruction or rapid spatial sampling~\citep{Chen2023a,Chen2023b,Zhao2026,Mi2026Leidenfrost}.
These approaches enable quantitative vapor-field measurements, but remain inherently sequential, with spatial resolution limited by beam size and scan spacing and temporal resolution by the time required to scan the measurement domain.

Infrared absorption imaging provides a direct route to combine the species selectivity and quantitative nature of absorption spectroscopy with spatially resolved imaging.
In contrast to scanning-based absorption measurements, the entire field of view is recorded simultaneously on an imaging detector, enabling instantaneous two-dimensional measurements without mechanical beam scanning.
Camera-based absorption imaging in the ultraviolet and visible spectral ranges has enabled quantitative imaging of ions, atoms, and their compounds through electronic transitions~\citep{Thorin2023,Thorin2025,Wang2026AlO}. In the mid-wave infrared, molecular rovibrational transitions have been exploited to image hydrocarbons, CO, CO$_2$, and temperature, progressing from two-dimensional measurements to time-resolved and volumetric reconstructions~\citep{Wei2018LAI,Schwarm2019LAI,Wei2024LAI}.
When the absorption band of the target species lies within the spectral response of the imaging system, the transmitted intensity can be converted through the Beer--Lambert relation into a spatial map of absorbance and, after calibration, species concentration.
The technique is therefore inherently tracer-free, species-selective, and quantitative, while retaining the parallel acquisition capability of an imaging system.
These features are particularly attractive for evaporating flows, for which introducing fluorescent tracers into the vapor phase may be difficult and steep concentration variations can occur close to the liquid--gas interface.

Nevertheless, before applying this imaging method to more complex evaporation phenomena, its quantitative measurement capability must first be assessed against a transport problem with a well-defined theoretical solution.
The one-dimensional Stefan evaporation problem provides a stringent benchmark for this purpose.
When an evaporating species diffuses through a stagnant carrier gas, the condition of zero carrier-gas molar flux produces a nonlinear vapor-concentration profile, rather than the linear profile predicted by the dilute Fickian approximation.
This problem therefore provides several independent theoretical features against which the experimental measurements can be evaluated.

In the present work, a vertically oriented Hele--Shaw cell is employed to establish a nearly one-dimensional vapor-transport region above an evaporating liquid.
Mid-infrared absorption images are converted into two-dimensional vapor mole-fraction fields using a spectrally weighted absorption coefficient, and the measurement method is evaluated through a hierarchy of increasingly stringent tests.
First, the spectroscopic conversion is assessed to establish the absolute concentration scale.
Second, the transverse uniformity of the two-dimensional concentration field is quantified to examine the validity of the one-dimensional approximation within the central measurement region.
Complete dimensional concentration profiles obtained at different gas-layer thicknesses and for different hydrocarbon liquids are then compared with the Stefan solution using only the experimentally measured endpoint conditions, without fitting the interior concentration data, thereby testing the spatial accuracy of the measurement.
Finally, normalized and logarithmically transformed concentration profiles are used to examine the predicted similarity and functional structure independently of the dimensional length scale, and thus to determine whether the measurements reproduce the nonlinear structure of Stefan transport.

This hierarchical, physics-based validation establishes the quantitative capability of infrared absorption imaging and provides a foundation for its future application to more complex multidimensional and unsteady evaporating flows.
Such measurements are particularly relevant to multicomponent droplets and liquid films, where evaporation couples heat and mass transport to fluid motion, phase separation, and thermal--solutal instabilities~\citep{Wakata2026EIPS,Wakata2025}, and to confined mass-transfer systems in which density-driven convection reorganizes the concentration field and alters the overall transport~\citep{Li2025Darcy}.
Although the vapor species examined in the present work are alkanes commonly employed in studies of Stefan evaporation, the applicability of the method is not restricted to these species.
With an appropriate spectral band and species-specific calibration, the same measurement principle can in principle be extended to other infrared-active vapors, including water and alcohol vapors.

\section{Measurement principle and transport model}
\label{sec:model}

\subsection{Geometry and measured quantity}\label{subsec:geometry}

\begin{figure*}[tb]
\centering
\includegraphics[width=0.9\textwidth]{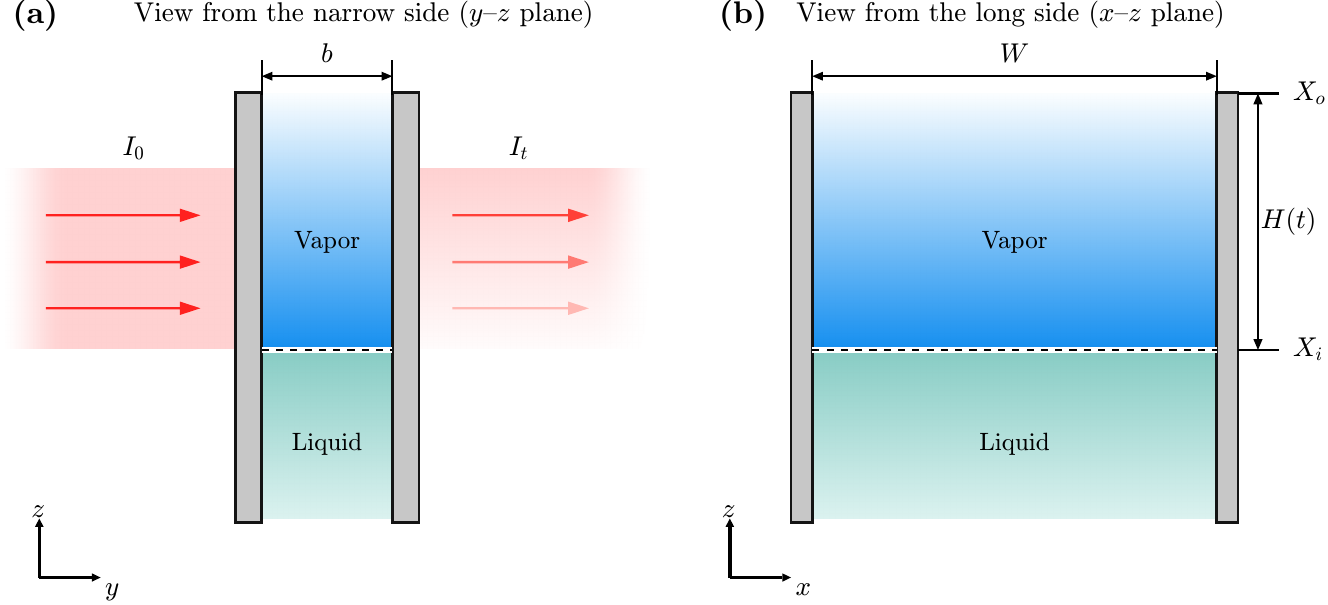}
\caption{Idealized measurement geometry. Infrared attenuation is recorded across the narrow
optical gap $b$ while the axial vapor profile develops between the liquid interface and the
opening.}
\label{fig:schematic}
\end{figure*}

The working geometry is sketched in Fig.~\ref{fig:schematic}. An evaporating liquid occupies the lower part of a vertical Hele--Shaw cell. The coordinate $z$ is measured vertically from the liquid--gas interface at $z=0$ toward the open end of the cell at $z=H$. The $y$ coordinate is directed across the narrow gap of the Hele--Shaw cell and coincides with the optical path, whereas $x$ denotes the transverse direction perpendicular to the optical path. The plate spacing, which also defines the nominal absorption path length, is denoted by $b$, and the cell width by $W$. The local vapor mole fraction is denoted by $X(x,y,z,t)$.

The infrared camera does not measure the vapor mole fraction at a mathematical point along the $y$ direction. Instead, the recorded signal results from the attenuation accumulated along the entire optical path through the cell. The illumination wavelength is selected such that it is absorbed by the vapor species of interest while the surrounding air is effectively transparent. The reference intensity recorded in the absence of vapor is denoted by $I_0(x,z)$, whereas the instantaneous transmitted intensity in the presence of vapor is denoted by $I_t(x,z,t)$. Owing to vapor absorption, $I_t<I_0$, and the attenuation is quantified through the transmittance and absorbance,
\begin{equation}
\mathcal{T}(x,z,t)=\frac{I_t(x,z,t)}{I_0(x,z)},
\qquad
\mathcal{A}(x,z,t)=-\ln \mathcal{T}(x,z,t).
\label{eq:absorbance}
\end{equation}

According to the Beer--Lambert law, the measured absorbance is related to the line-integrated vapor mole fraction through
\begin{equation}
\mathcal{A}(x,z,t)=\int_{-b/2}^{b/2}\kappa X(x,y,z,t)\dd y,
\label{eq:beer-lambert}
\end{equation}
where $\kappa$ is an effective absorption coefficient evaluated for the experimental temperature and pressure, accounting for the spectral overlap between the vapor absorption band and the illumination spectrum. Its determination from spectroscopic calculations and its independent experimental validation are described in Sect.~\ref{subsec:kappa clibration}. When the thermodynamic conditions and optical properties may be regarded as uniform across the narrow gap, $\kappa$ is independent of $y$, and equation~\eqref{eq:beer-lambert} reduces to
\begin{equation}
\mathcal{A}(x,z,t)=\kappa b\overline{X}(x,z,t),
\label{eq:beer-lambert-avg}
\end{equation}
where
\begin{equation}
\overline{X}(x,z,t)=\frac{1}{b}\int_{-b/2}^{b/2}X(x,y,z,t)\dd y
\label{eq:gap-average}
\end{equation}
is the gap-averaged vapor mole fraction. Thus, once $\kappa$ has been established for the experimental conditions, the measured absorbance provides a direct quantitative measure of the vapor mole fraction averaged across the Hele--Shaw gap.

The geometry and boundary conditions further permit the measured concentration field to be treated as effectively one-dimensional over the region of interest. Across the narrow gap, the characteristic diffusion time scales as $b^2/D$, which is much shorter than the corresponding diffusion time $H^2/D$ over the vapor layer because $b\ll H$. Concentration variations in the $y$ direction are therefore rapidly homogenized by molecular diffusion, so that the gap-averaged quantity $\overline{X}(x,z,t)$ closely represents the local vapor mole fraction within the gap.

Along the $x$ direction, the liquid interface, the cell geometry, and the boundary conditions are approximately uniform over the central measurement region. Away from the lateral sidewalls, no appreciable concentration gradient is therefore expected parallel to the interface, which is also validated by experiments described in Sect.~\ref{subsec:1D validation}. The dominant concentration variation is instead imposed in the $z$ direction by the contrast between the vapor-rich liquid interface at $z=0$ and the vapor-depleted opening at $z=H$. Consequently, within the measurement region considered here,
\begin{equation}
X(x,y,z,t)
\simeq
\overline{X}(x,z,t)
\simeq
X(z,t),
\label{eq:one-dimensional}
\end{equation}
and the vapor transport is henceforth described in terms of the one-dimensional mole-fraction field $X(z,t)$.

\subsection{One-dimensional gas-phase transport model}\label{subsec:transport-model}

Gas-phase transport is considered over $0\le z\le H$, with the vapor and carrier gas designated as species $A$ and $B$, respectively. The mixture is treated as an ideal binary gas at uniform temperature $T$ and pressure $P$, giving the total molar concentration
\begin{equation}
C_T=\frac{P}{RT},
\label{eq:total-molar-concentration}
\end{equation}
where $R$ is the universal gas constant. The binary diffusion coefficient $D\equiv D_{AB}$ is taken as constant, and homogeneous chemical reactions in the gas phase are neglected.

The characteristic diffusion time across the gas layer and the characteristic time associated with interface motion are estimated as
\begin{equation}
\tau_D\sim\frac{H^2}{D_{AB}},
\qquad
\tau_m\sim\frac{H}{|\dot H|}.
\end{equation}

For the present experiments,
$H\sim O(10^{-2})~\mathrm{m}$,
$|\dot H|\sim O(10^{-5})~\mathrm{m\,s^{-1}}$, 
$D_{AB}\sim O(10^{-5})~\mathrm{m^2\,s^{-1}}$, and hence
\begin{equation}
\frac{\tau_D}{\tau_m}
\sim\frac{H|\dot H|}{D_{AB}}
\sim O(10^{-2})\ll1.
\end{equation}

The gas-phase concentration field is therefore treated as quasi-steady. At each instant, its spatial profile adjusts to the instantaneous gas-layer height $H(t)$ and endpoint mole fractions, whose slower temporal evolution is retained parametrically.

The endpoint mole fractions are defined as
\begin{equation}
X_i(t)\equiv X(0,t),
\qquad
X_o(t)\equiv X(H,t).
\label{eq:generic-endpoint-values}
\end{equation}

Their physical boundary conditions are discussed below. For clarity, the explicit dependence on $t$ is omitted in the following derivation.

\subsubsection{Stefan transport}

Let $N_A$ and $N_B$ denote the molar fluxes of vapor and carrier gas, respectively, in the positive $z$ direction, and let $N=N_A+N_B$ denote the total molar flux. For a one-dimensional binary mixture, the vapor flux is
\begin{equation}
N_A=XN-C_TD\frac{\dd X}{\dd z},
\label{eq:general-vapor-flux}
\end{equation}
where the two terms on the right-hand side represent transport by bulk molar motion and molecular diffusion, respectively. Under source-free quasi-steady conditions, species conservation gives $\dd N_A/\dd z=0$, so $N_A$ is spatially uniform.

For evaporation into a non-condensing carrier gas, the zero carrier-gas flux at the liquid interface and the absence of sources or sinks within the gas layer imply $N_B=0$ throughout the gas phase~\citep{Bird2002,Mitrovic2012}. Consequently, $N=N_A$, and Eq.~\eqref{eq:general-vapor-flux} becomes
\begin{equation}
N_A=-\frac{C_TD}{1-X}\frac{\dd X}{\dd z}.
\label{eq:stefan-flux-local}
\end{equation}

Integration between $z=0$ and $z=H$, subject to $X(0)=X_i$ and $X(H)=X_o$, gives
\begin{equation}
\ln(1-X)
=\left(1-\frac{z}{H}\right)\ln(1-X_i)
+\frac{z}{H}\ln(1-X_o).
\label{eq:stefan-log-linear}
\end{equation}
The corresponding Stefan mole-fraction profile is
\begin{equation}
X_S(z)=1-(1-X_i)^{1-z/H}(1-X_o)^{z/H}.
\label{eq:stefan-profile}
\end{equation}

In terms of the normalized coordinate $\eta=z/H$ and the normalized mole fraction $X^*=(X-X_o)/(X_i-X_o)$, the normalized Stefan profile becomes
\begin{equation}
X_S^*(\eta)
=
\frac{1-(1-X_i)^{1-\eta}(1-X_o)^{\eta}-X_o}
{X_i-X_o}.
\label{eq:stefan-normalized}
\end{equation}

The logarithmic form in Eq.~\eqref{eq:stefan-log-linear} provides a complementary parameter-free representation of the Stefan profile. Defining the Stefan-transformed concentration as
\begin{equation}
Y_{\mathrm{S}}
\equiv
\frac{\ln[(1-X)/(1-X_i)]}
{\ln[(1-X_o)/(1-X_i)]},
\qquad
Y_{\mathrm{S}}=\eta,
\label{eq:stefan-transformed}
\end{equation}
the exact Stefan solution reduces to an identity relation that is independent of $H$, $X_i$, and $X_o$.

In the dilute-vapor limit, $X\ll1$, the finite-concentration correction associated with Stefan transport becomes negligible. Using $\ln(1-X)\simeq-X$ in Eq.~\eqref{eq:stefan-log-linear} gives
\begin{equation}
X_L(z)=X_i+(X_o-X_i)\frac{z}{H},
\qquad
X_L^*(\eta)=1-\eta.
\label{eq:linear-diffusion-normalized}
\end{equation}
Thus, the linear-diffusion profile is recovered as the dilute limit of the Stefan solution.

\subsubsection{Boundary conditions}

The endpoint mole fractions are determined by the conditions at the liquid surface and the cell opening. For a pure liquid under local thermodynamic equilibrium,
\begin{equation}
X_i=X_{\mathrm{sat}}=\frac{p_{\mathrm{sat}}(T_i)}{P},
\label{eq:equilibrium-interface-bc}
\end{equation}
where $T_i$ is the interfacial temperature and $P$ is the total pressure~\citep{Mitrovic2012,Vlasov2020}. At the cell opening, $X_o$ is governed by vapor removal and need not vanish when the external sweep does not completely remove the vapor.

In the comparisons below, $X_i$ and $X_o$ are taken directly from the measured endpoints and applied identically to the Stefan solution and its linear-diffusion limit. No idealized endpoint concentration is imposed, so the comparison tests the predicted concentration distribution within the gas layer. Because the source-free Stefan relation holds over any subinterval of the gas layer, the same comparison remains valid after redefining $z$ and $H$ relative to the retained interval endpoints.

\section{Experimental setup}
\label{sec:methods}

The vertically mounted Hele--Shaw cell consisted of two sapphire windows clamped against a metal spacer, which defined a nominal gap of $b=0.8~\mathrm{mm}$ and a lateral width of $W=80~\mathrm{mm}$. The lower portion of the cell was filled with a liquid hydrocarbon, forming a nominally planar liquid--gas interface and a vapor layer of instantaneous height $H(t)$ between the interface and the opening, with the coordinate system defined in Fig.~\ref{fig:schematic}. Quantitative measurements were restricted to the central region of the cell to minimize sidewall and edge effects, and the optical surfaces were cleaned before each experiment.

Table~\ref{tab:fluid-properties} summarizes the thermophysical properties of hydrocarbons used in the analysis. The saturation pressures at $298.15~\mathrm{K}$ were obtained from the NIST Chemistry WebBook~\citep{NISTPentane,NISTHexane,NISTHeptane} and converted to $X_{\mathrm{sat}}$ using $P=101.325~\mathrm{kPa}$, while the binary diffusion coefficients in nitrogen $D_{A,\mathrm{N_2}}$ were estimated at $298.15~\mathrm{K}$ and $1~\mathrm{atm}$ using the Fuller--Schettler--Giddings correlation~\citep{Fuller1966}.

\begin{table}[htb]
\centering
\caption{Thermophysical properties of the hydrocarbon working fluids at
$T=298.15~\mathrm{K}$ and $P=1~\mathrm{atm}$.}
\label{tab:fluid-properties}
\begin{tabular}{lcccc}
\hline
Fluid
& Formula
& $p_{\mathrm{sat}}$
& $X_{\mathrm{sat}}$
& $D_{A,\mathrm{N_2}}$
\\
& & $(\mathrm{kPa})$ & & $(10^{-6}~\mathrm{m^2\,s^{-1}})$ \\
\hline
\textit{n}-Pentane & $\mathrm{C_5H_{12}}$ & 68.36 & 0.6746 & 8.81 \\
\textit{n}-Hexane  & $\mathrm{C_6H_{14}}$ & 20.16 & 0.1990 & 7.98 \\
\textit{n}-Heptane & $\mathrm{C_7H_{16}}$ & 6.100 & 0.0602 & 7.34 \\
\hline
\end{tabular}
\end{table}

The vapor outlet of the Hele--Shaw cell was connected to a 3D-printed purge chamber. A controlled flow of dry nitrogen passed through porous flow-conditioning elements before sweeping across the cell opening, continuously removing vapor transported out of the gap. This arrangement stabilized the upper boundary condition while minimizing disturbance to transport within the Hele--Shaw cell. The vapor mole fraction at the opening was not assumed to vanish. Instead, its measured value was used directly in the subsequent theoretical comparisons.

Infrared absorption imaging was used to measure the vapor concentration field inside the Hele--Shaw cell. Broadband infrared illumination from a silicon-carbide Globar source, with an output wavelength range of $500$--$9000~\mathrm{nm}$, was collimated and expanded using reflective optics to provide approximately uniform irradiation over the measurement region. After traversing the cell along the $y$ direction, the transmitted radiation was filtered by a narrow-band optical filter mounted inside the camera and then recorded by a mid-infrared camera. The filter had a central wavelength of $3420~\mathrm{nm}$ and a FWHM of $75~\mathrm{nm}$, selecting a spectral region in which hydrocarbon vapor exhibits appreciable absorption while the background gas remains effectively transparent. The resulting images provided the spatially resolved transmitted intensity $I_t(x,z,t)$ with a calibrated object-plane sampling of $30.5~\mu\mathrm{m\,pixel^{-1}}$, an exposure time of $2500~\mu\mathrm{s}$, and a frame rate of $1~\mathrm{Hz}$.

Reference images required for the absorption measurement were first recorded under the corresponding vapor-free and background conditions. The evaporation experiment was then initiated, and infrared images were acquired continuously as the liquid interface receded and the vapor concentration field evolved. During the analysis, the instantaneous interface position defined the origin of the moving coordinate system. Accordingly, $z=0$ corresponded to the liquid--gas interface and $z=H(t)$ to the upper boundary of the measured gas layer.

\section{Measurement characterization and validation}

\subsection{Calibration of spectrally weighted absorption coefficient}\label{subsec:kappa clibration}

The species-specific effective absorption coefficient $\kappa_j$ required in Eq.~\eqref{eq:beer-lambert-avg} was first estimated spectroscopically and then determined independently from saturated-vapor measurements. The spectroscopic calculation combines the vapor absorption spectrum with the filter transmission to predict the band-integrated coefficient, while the equilibrium measurement provides the coefficient used to convert the measured absorbance fields into vapor mole fraction.

For vapor species $j$, the wavelength-dependent transmission through the optical path $b$ is
\begin{equation}
\mathcal{T}_{\lambda,j}(X)
=
\exp\left[-\kappa_{\lambda,j}bX\right],
\label{eq:spectral-transmission}
\end{equation}
where $\kappa_{\lambda,j}$ is the wavelength-dependent coefficient at the experimental temperature and pressure. Because the $75~\mathrm{nm}$ filter bandwidth is narrow relative to the spectral ranges of the source and detector, variations in source emission and detector responsivity across the passband were neglected. The weighting was therefore represented by the filter transmission $T_f(\lambda)$, giving
\begin{equation}
\begin{aligned}
\mathcal{T}_{\mathrm{eff},j}(X)
&=
\frac{\displaystyle\int T_f(\lambda)\mathcal{T}_{\lambda,j}(X)\dd\lambda}
{\displaystyle\int T_f(\lambda)\dd\lambda},\\
\mathcal{A}_{\mathrm{eff},j}(X)
&=-\ln\mathcal{T}_{\mathrm{eff},j}(X).
\end{aligned}
\label{eq:broadband-transmission}
\end{equation}

The calculations were performed over $X=0.01$--$0.80$ at $T=298.15~\mathrm{K}$, $P=1~\mathrm{atm}$, and $b=0.08~\mathrm{cm}$ using the corresponding hydrocarbon spectra~\citep{Goldenstein2017SpectraPlot,Gordon2022HITRAN}. Over this range, the band-integrated absorbance was represented by
\begin{equation}
\mathcal{A}_{\mathrm{eff},j}(X) = \kappa_{\mathrm{spec},j}bX,
\label{eq:spectroscopic-calibration}
\end{equation}
where $\kappa_{\mathrm{spec},j}$ was obtained from an origin-constrained linear fit.

\begin{figure}[t]
    \centering
    \includegraphics[width=0.9\columnwidth]{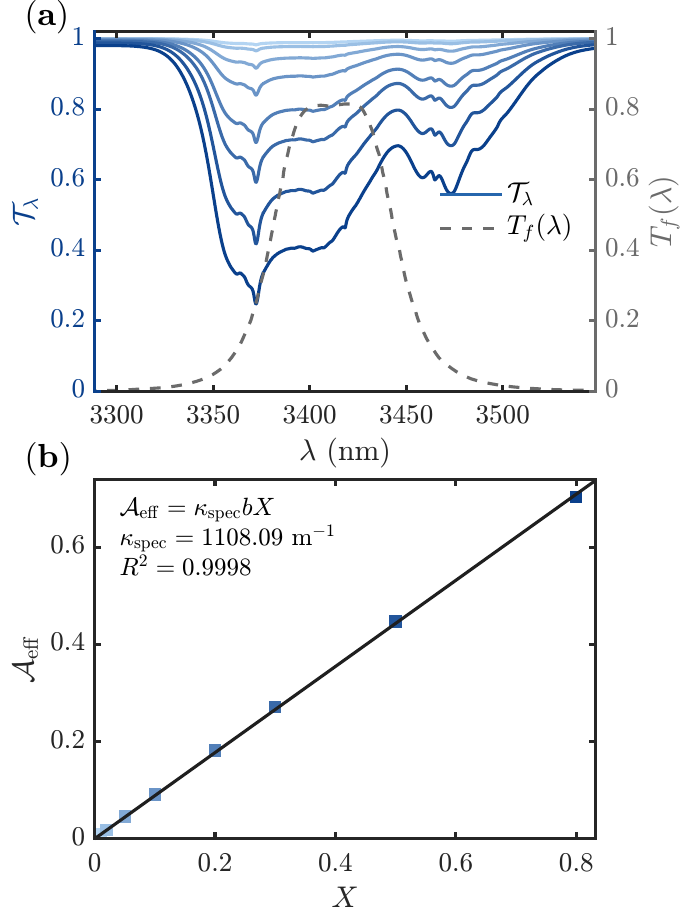}
    \caption{
    Spectroscopic calibration of the vapor mole-fraction measurement.
    (a) Calculated \textit{n}-pentane transmittance spectra for $X=0.01$--$0.80$ and the transmission function of the NB-3420-075-nm filter.
    Increasing color intensity denotes increasing vapor mole fraction.
    (b) Corresponding band-integrated absorbance.
    Squares denote the calculated values and the solid line is the origin-constrained linear fit.
    }
    \label{fig:spectroscopic-calibration}
\end{figure}

Figure~\ref{fig:spectroscopic-calibration} illustrates the calculation for \textit{n}-pentane. The transmission decreases across the hydrocarbon absorption band as $X$ increases, and the filter selects the spectral interval contributing to the recorded intensity. The origin-constrained fit gave $\kappa_{\mathrm{spec},p}=1108.09~\mathrm{m^{-1}}$ with $R^2=0.9998$. Applying the same procedure to \textit{n}-hexane and \textit{n}-heptane gave $\kappa_{\mathrm{spec},hx}=1322.91~\mathrm{m^{-1}}$ with $R^2=0.9995$ and $\kappa_{\mathrm{spec},hp}=1533.38~\mathrm{m^{-1}}$ with $R^2=0.9989$. The subscripts $p$, $hx$, and $hp$ denote \textit{n}-pentane, \textit{n}-hexane, and \textit{n}-heptane, respectively. Only the \textit{n}-pentane calculation is shown because the procedure is identical for the three species.

The spectroscopic estimates were then assessed through independent equilibrium measurements. For each liquid $j$, the calibration cell retained a small opening to the surroundings so that the total pressure remained equal to the ambient pressure while the selected measurement region approached vapor--liquid equilibrium. At steady state, the absorbance in the spatially uniform gas region is denoted by $\mathcal{A}_{\mathrm{sat},j}$, and the equilibrium vapor mole fraction is
\begin{equation}
X_{\mathrm{sat},j}
=
\frac{p_{\mathrm{sat},j}(T_{\mathrm{eq},j})}
{P_{\mathrm{eq},j}},
\label{eq:equilibrium-vapor-mole-fraction}
\end{equation}
where $T_{\mathrm{eq},j}$ and $P_{\mathrm{eq},j}$ are the equilibrium temperature and total pressure, respectively. The calibration temperature and pressure were $T_{\mathrm{eq},j}=298~\mathrm{K}$ and $P_{\mathrm{eq},j}=101~\mathrm{kPa}$ for all three species. The experimentally determined effective absorption coefficient is then
\begin{equation}
\kappa_{\mathrm{exp},j}
=
\frac{\mathcal{A}_{\mathrm{sat},j}}
{bX_{\mathrm{sat},j}}.
\label{eq:experimental-absorption-coefficient}
\end{equation}

\begin{figure}[htb]
    \centering
    \includegraphics[width=0.9\columnwidth]{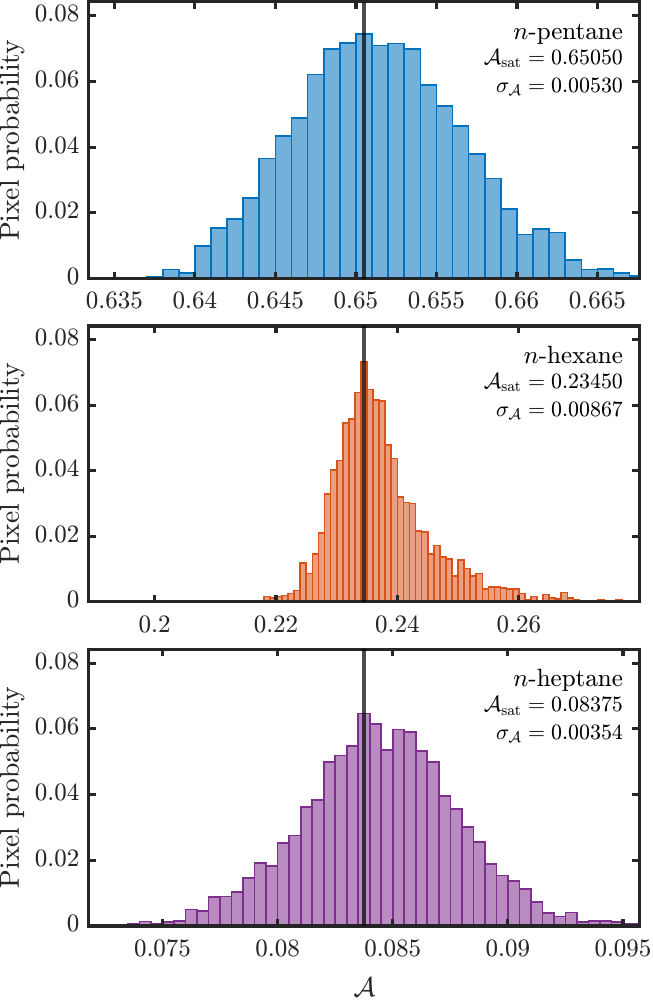}
    \caption{Pixelwise absorbance distributions within the selected spatially uniform saturated-vapor regions for \textit{n}-pentane, \textit{n}-hexane, and \textit{n}-heptane, shown from top to bottom. The solid vertical line marks the histogram peak adopted as $\mathcal{A}_{\mathrm{sat},j}$.}
    \label{fig:equilibrium-calibration-histograms}
\end{figure}

For each species, the peak of the pixelwise absorbance distribution in the selected uniform region was used as $\mathcal{A}_{\mathrm{sat},j}$, reducing sensitivity to sparse pixels in the distribution tails. The resulting histograms are shown in Fig.~\ref{fig:equilibrium-calibration-histograms}.

Together with the saturation pressures in Table~\ref{tab:fluid-properties}, the measured absorbances yielded $\kappa_{\mathrm{exp},p}=1205.24~\mathrm{m^{-1}}$, $\kappa_{\mathrm{exp},hx}=1473.26~\mathrm{m^{-1}}$, and $\kappa_{\mathrm{exp},hp}=1738.93~\mathrm{m^{-1}}$. These values exceed the corresponding spectroscopic estimates by $8.77\%$, $11.37\%$, and $13.41\%$. The consistently positive differences may reflect uncertainties in the database spectra, deviations of the actual source--filter--detector response from the simplified weighting model, and experimental uncertainties in the equilibrium conditions, optical path length, and absorbance extraction.

The concentration fields reported below were obtained from Eq.~\eqref{eq:beer-lambert-avg} using the independently determined $\kappa_{\mathrm{exp},j}$ for each species. Following the one-dimensional reduction established above, the corresponding profiles are expressed as $X_j(z,t)$.

\subsection{Validation of the one-dimensional approximation}\label{subsec:1D validation}

The one-dimensional reduction in Eq.~\eqref{eq:one-dimensional} requires concentration variations across both $x$ and $y$ to remain small relative to the variation along $z$. Uniformity across the narrow $y$ direction is supported by the short transverse diffusion time associated with the Hele--Shaw gap. The remaining assumption, namely that the concentration is approximately uniform along $x$ over the analyzed region, can be evaluated directly from the measured two-dimensional field.

\begin{figure}[t]
    \centering
    \includegraphics[width=0.9\columnwidth]{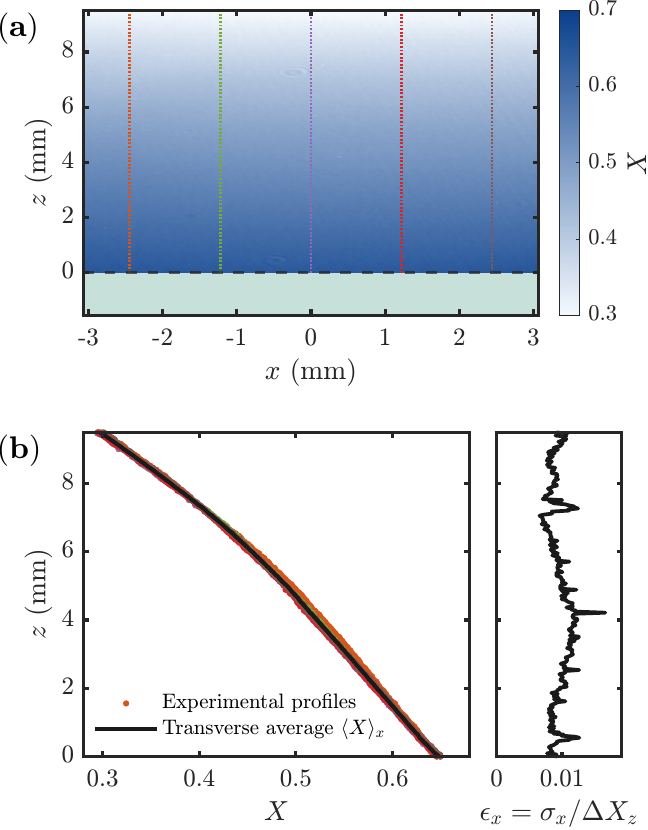}
    \caption{
    Validation of the one-dimensional approximation.
    (a) Instantaneous two-dimensional vapor mole-fraction field $X(x,z)$.
    The dashed horizontal line marks the liquid--gas interface, and the colored vertical lines indicate the profile locations.
    (b) Profiles at the five transverse positions and their transverse mean $\langle X\rangle_x$.
    The adjoining narrow panel shows the normalized transverse non-uniformity $\epsilon_x$.
    }
    \label{fig:1D validation by experiments}
\end{figure}

Figure~\ref{fig:1D validation by experiments}(a) presents a representative \textit{n}-pentane mole-fraction field. Quantitative analysis was restricted to the central portion of the field of view to reduce possible contributions from the sidewalls and optical edge effects. The five colored vertical lines mark the transverse locations used to extract the profiles shown in Fig.~\ref{fig:1D validation by experiments}(b). At each location, the profile was averaged over an 11-pixel-wide window rather than taken from a single pixel column, reducing pixel-scale noise and the influence of isolated defective pixels.

The residual transverse variation was quantified relative to the concentration change along the principal transport direction. The transverse mean, standard deviation, and normalized non-uniformity were defined as
\begin{equation}
\begin{aligned}
\left\langle X\right\rangle_x(z,t)
&=\frac{1}{L_x}\int_{x_1}^{x_2}X(x,z,t)\dd x,\\
\sigma_x^2(z,t)
&=\frac{1}{L_x}\int_{x_1}^{x_2}
\left[X(x,z,t)-\left\langle X\right\rangle_x\right]^2\dd x,\\
\epsilon_x(z,t)
&=\frac{\sigma_x(z,t)}{\Delta X_z(t)},
\qquad
\Delta X_z=X_i-X_o,
\end{aligned}
\label{eq:transverse-nonuniformity}
\end{equation}
where $L_x=x_2-x_1$ is the width of the selected region and $\Delta X_z$ is the mole-fraction change along the principal transport direction. Normalization by $\Delta X_z$ expresses the transverse variation relative to the axial concentration difference and allows its magnitude to be interpreted independently of the absolute vapor loading.

The concentration field in Fig.~\ref{fig:1D validation by experiments}(a) exhibits approximately horizontal iso-concentration contours over the selected region. The concentration decreases primarily with increasing $z$, while no systematic displacement or curvature of the contours is observed across the five profile locations. Localized image features remain visible at a few positions, but they do not form a persistent transverse concentration gradient over the central region.

Consistently, the five profiles in Fig.~\ref{fig:1D validation by experiments}(b) nearly collapse onto the transverse mean throughout the analyzed gas-layer interval. Their separation is much smaller than the overall concentration change between the lower and upper ends of the interval. The adjoining narrow panel shows that $\epsilon_x$ remains close to $0.01$ over most of the gas layer, apart from localized excursions associated with image non-uniformity. Thus, the transverse standard deviation is typically about one percent of the axial mole-fraction difference.

These observations show that the dominant concentration gradient is directed along $z$ within the selected measurement region. The measured field can therefore be represented by the transverse average $\left\langle X\right\rangle_x(z,t)$ for comparison with the one-dimensional Stefan model. All concentration profiles examined below are obtained using this average.
\begin{figure*}[t]
    \centering
    \includegraphics[width=\textwidth]{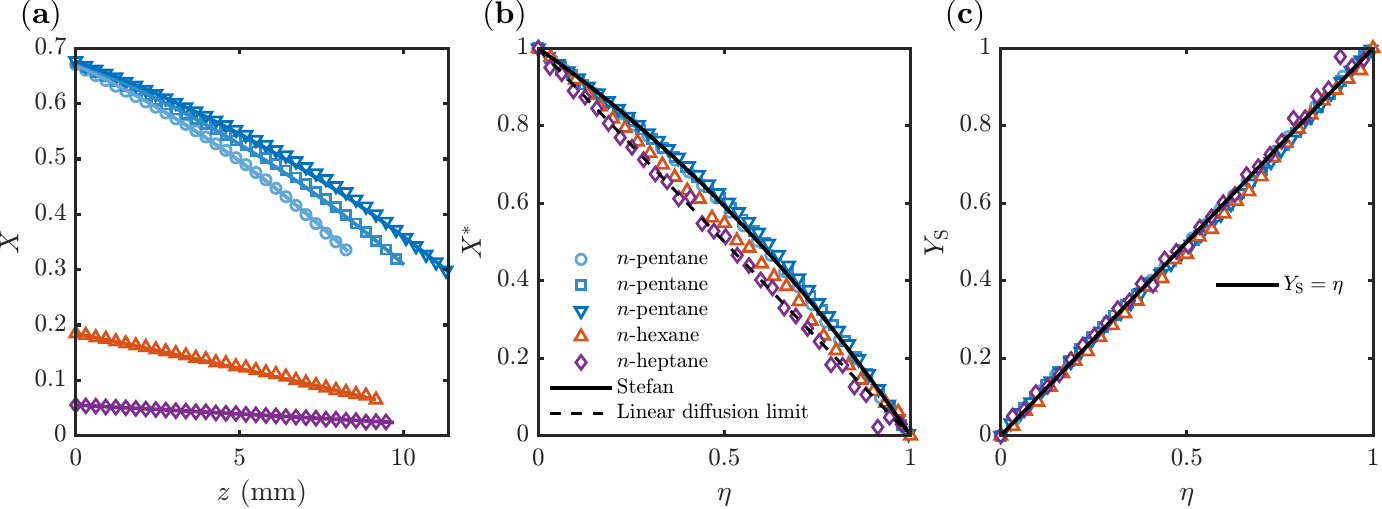}
    \caption{
    Experimental validation of infrared vapor-concentration measurements against the one-dimensional Stefan solution.
    (a) Dimensional mole-fraction profiles. Symbols denote the measurements, and the colored solid curves denote the corresponding Stefan predictions.
    (b) Profiles expressed in the normalized coordinates $(\eta,X^*)$. The solid black curve denotes a representative Stefan profile, and the dashed line denotes the linear-diffusion limit.
    (c) Stefan-transformed profiles expressed in terms of $(\eta,Y_{\mathrm{S}})$. The solid black line denotes the identity relation $Y_{\mathrm{S}}=\eta$ predicted by the Stefan solution.
    }
    \label{fig:profile_similarity}
\end{figure*}

\subsection{Validation of infrared concentration measurements against Stefan transport}\label{sec:profile_similarity}

The classical one-dimensional Stefan problem is used here as a quantitative benchmark for infrared absorption imaging. The objective is not to identify a new evaporation mechanism, but to assess whether the measured concentration field reproduces a known transport solution. The comparison was performed over a retained gas-phase interval that excluded the optically mixed region adjacent to the liquid--gas interface. Here, $z=0$ and $z=H$ denote the lower and upper ends of this interval, and $X_i$ and $X_o$ are the corresponding measured endpoint concentrations.

Figure~\ref{fig:profile_similarity} compares three \textit{n}-pentane profiles at $H=8.31$, $10.02$, and $11.36~\mathrm{mm}$ with one \textit{n}-hexane profile at $H=9.16~\mathrm{mm}$ and one \textit{n}-heptane profile at $H=9.71~\mathrm{mm}$. For each dataset, the measured $H$, $X_i$, and $X_o$ were supplied directly to the theoretical solution without fitting the interior concentration data, which therefore provide an independent test of the predicted spatial distribution.

The dimensional comparison in Fig.~\ref{fig:profile_similarity}(a) spans concentration fields that differ appreciably in both magnitude and shape. The \textit{n}-pentane profiles have the largest lower-end mole fractions and exhibit the clearest nonlinear curvature, while the lower-loading \textit{n}-hexane and \textit{n}-heptane profiles are progressively closer to linear. The three \textit{n}-pentane profiles also represent different stages of interface recession. As the interface recedes, $H$ increases and the measured outlet mole fraction $X_o$ decreases, so the later profiles cannot be obtained by simply stretching the $z$ coordinate. Despite these differences, the Stefan predictions follow all five measured distributions throughout the resolved gas layer. The NRMSE between the measured interior concentrations and the Stefan predictions, normalized by $X_i-X_o$, ranged from $0.0055$ to $0.0185$, corresponding to deviations of only $0.55\%$--$1.85\%$ of the measured axial concentration difference.

Figure~\ref{fig:profile_similarity}(b) provides a more sensitive test of profile shape by removing the direct effects of the dimensional layer thickness and concentration difference. The three normalized \textit{n}-pentane profiles nearly collapse onto the representative nonlinear Stefan curve, which was evaluated using the mean value of $\ln[(1-X_o)/(1-X_i)]$ for these profiles. Because the Stefan and linear-diffusion predictions satisfy the same endpoint concentrations, their difference arises entirely from the interior concentration distribution. For the three \textit{n}-pentane profiles, the Stefan NRMSE values were $0.0055$, $0.0093$, and $0.0124$, compared with $0.0647$, $0.0739$, and $0.0809$ for the linear-diffusion limit. The corresponding error reductions of $91.4\%$, $87.4\%$, and $84.6\%$ demonstrate that the measurements resolve the finite-concentration curvature rather than merely recovering the overall concentration difference. At the lower vapor loadings of \textit{n}-hexane and \textit{n}-heptane, the normalized profiles progressively approach the linear-diffusion limit, as expected from Eq.~\eqref{eq:linear-diffusion-normalized}. The Stefan and linear-limit NRMSE values are $0.0185$ and $0.0302$ for \textit{n}-hexane, and $0.0145$ and $0.0137$ for \textit{n}-heptane, respectively. At the \textit{n}-heptane loading, the Stefan and linear predictions are already nearly indistinguishable and therefore yield comparable errors. The measurements therefore reproduce both expected regimes, namely resolvable finite-concentration curvature at high vapor loading and an approximately linear profile as the finite-concentration correction weakens. This transition also indicates that the curvature observed for \textit{n}-pentane is not an optical or processing feature imposed on every dataset.

A complementary parameter-free representation is provided by the logarithmic transformation in Eq.~\eqref{eq:stefan-transformed}. In Fig.~\ref{fig:profile_similarity}(c), normalization by the measured endpoints removes the differences in $H$, $X_i$, and $X_o$, leaving the identity relation predicted by the Stefan solution. Although the five datasets differ strongly in concentration level and in the apparent curvature of $X(z)$, their interior measurements collapse close to this line. The endpoints map to zero and unity by construction, so the interior points constitute the relevant test. Linear fits to the transformed interior data give slopes of $0.9803$--$1.0107$, intercepts of $-0.0131$--$0.0078$, and $R^2$ values of $0.9984$--$0.9997$. The mean absolute deviations from Eq.~\eqref{eq:stefan-transformed} are $0.0049$--$0.0159$.

The three representations isolate complementary aspects of the measurement. The dimensional profiles test the recovered concentration magnitude and spatial variation, the conventional normalization distinguishes finite-concentration curvature from the dilute limit, and the logarithmic transformation tests the functional structure of the Stefan solution across different liquids and interval lengths. Together with the independent concentration calibration in Sect.~\ref{subsec:kappa clibration} and the transverse-uniformity assessment in Sect.~\ref{subsec:1D validation}, these comparisons show that the imaging method recovers both the absolute concentration scale and the predicted spatial structure without fitting the interior profile. Within the tested concentration and geometric ranges, this agreement supports application of the method to less canonical evaporating flows for which the concentration field is not known a priori.

\section{Conclusions}

This work developed and quantitatively evaluated an infrared absorption imaging method for measuring spatially resolved vapor concentration fields during evaporation.
Broadband mid-infrared attenuation was related to vapor mole fraction through a spectrally weighted absorption coefficient, enabling two-dimensional, tracer-free measurements in a vertically oriented Hele--Shaw cell with a calibrated object-plane sampling of $30.5~\mu\mathrm{m\,pixel^{-1}}$. The spectroscopic conversion was assessed against independent saturated-vapor measurements. The resulting experimental coefficients were $1205.24$, $1473.26$, and $1738.93~\mathrm{m^{-1}}$ for \textit{n}-pentane, \textit{n}-hexane, and \textit{n}-heptane, respectively. These values were $8.77\%$, $11.37\%$, and $13.41\%$ above the corresponding spectroscopic calculations. The consistently positive differences may reflect uncertainties in the database spectra, deviations of the actual source--filter--detector response from the simplified weighting model, and experimental uncertainties in the equilibrium conditions, optical path length, and absorbance extraction. The experimentally determined coefficients were used for the reported concentration fields.

The two-dimensional measurements showed that the concentration field was nearly uniform across the central portion of the cell, with a normalized transverse non-uniformity of approximately $0.01$ over the selected region.
This observation supports the one-dimensional representation used for comparison with the classical Stefan solution, while also defining the part of the field over which that approximation is justified.

Five concentration profiles spanning different retained gas-phase interval lengths and three hydrocarbon liquids were then tested against the Stefan prediction using the measured interval length and endpoint mole fractions, without fitting the interior concentration data.
The \textit{n}-pentane measurements resolved the finite-concentration curvature of the Stefan profile and were distinguishable from the linear diffusion limit, whereas the lower vapor loadings of \textit{n}-hexane and \textit{n}-heptane approached the dilute linear form.
Across the five measurements, the normalized root-mean-square deviations from the Stefan prediction were $0.0055$--$0.0185$.
After logarithmic Stefan transformation and normalization, the profiles collapsed onto the predicted identity relation, with fitted slopes of $0.9803$--$1.0107$, intercepts of $-0.0131$--$0.0078$, $R^2=0.9984$--$0.9997$, and mean absolute deviations of $0.0049$--$0.0159$.

Together, these results show that the imaging method recovers not only representative concentration levels but also the dimensional and normalized spatial structure predicted for a canonical evaporation problem.
The agreement is therefore used here as a validation of the diagnostic rather than as evidence for a new transport mechanism.
Within the tested range of vapor loading, gas-layer thickness, and cell geometry, infrared absorption imaging provides a quantitative full-field measurement of vapor concentration and its temporal evolution.
This capability establishes a basis for future measurements of multidimensional and unsteady evaporating flows, where point measurements and analytical concentration solutions are insufficient.

\section*{Data availability}

The data and analysis code that support the findings of this article are not currently publicly available because the repository is being prepared. They are available from the corresponding authors upon reasonable request and will be deposited in a public repository upon publication.

\section*{Author contributions}

Z.S.: conceptualization, methodology, investigation, software, formal analysis, data curation, visualization, and writing -- original draft. W.W.: methodology, validation, and writing -- review and editing. C.S.: supervision, resources, funding acquisition, and writing -- review and editing. X.C.: conceptualization, methodology, supervision, funding acquisition, and writing -- review and editing.

\begin{acknowledgments}
This work was supported by the National Natural Science Foundation of China (Nos.~52576181, 12588201, and 52506202), the New Cornerstone Science Foundation through the New Cornerstone Investigator Program and the XPLORER PRIZE, the China Postdoctoral Science Foundation (No.~2025M771854), and the Shuimu Tsinghua Scholar Program (No.~2024SM298).
\end{acknowledgments}

\section*{Declaration of interests}

The authors declare no competing interests.

\bibliography{references}

@article{Goldenstein2017SpectraPlot,
  author  = {Goldenstein, Christopher S. and Miller, Victor A. and
             Spearrin, R. Mitchell and Strand, Christopher L.},
  title   = {{SpectraPlot.com}: Integrated Spectroscopic Modeling of
             Atomic and Molecular Gases},
  journal = {J. Quant. Spectrosc. Radiat. Transfer},
  volume  = {200},
  pages   = {249--257},
  year    = {2017},
  doi     = {10.1016/j.jqsrt.2017.06.007}
}

@article{Gordon2022HITRAN,
  author  = {Gordon, Iouli E. and Rothman, Laurence S. and
             Hargreaves, Robert J. and others},
  title   = {The HITRAN2020 Molecular Spectroscopic Database},
  journal = {J. Quant. Spectrosc. Radiat. Transfer},
  volume  = {277},
  pages   = {107949},
  year    = {2022},
  doi     = {10.1016/j.jqsrt.2021.107949}
}

@article{Wang2019LAS,
  author  = {Wang, Zhenhai and Fu, Pengfei and Chao, Xing},
  title   = {Laser Absorption Sensing Systems: Challenges, Modeling, and Design Optimization},
  journal = {Appl. Sci.},
  volume  = {9},
  number  = {13},
  pages   = {2723},
  year    = {2019},
  doi     = {10.3390/app9132723}
}

@article{Zhu2022Comb,
  author  = {Zhu, Ning and Xu, Ze and Wang, Zhenhai and Song, Zihao and Wang, Weitian and Chen, Xiaoliang and Chao, Xing},
  title   = {Midinfrared Compressed Fourier-Transform Spectroscopy with an Optical Frequency Comb},
  journal = {Phys. Rev. Applied},
  volume  = {18},
  number  = {2},
  pages   = {024025},
  year    = {2022},
  doi     = {10.1103/PhysRevApplied.18.024025}
}

@article{Wang2025HPSDS,
  author  = {Wang, Weitian and Fu, Pengfei and Song, Zihao and Wang, Zhenhai and Zhu, Ning and Ma, Liuhao and Chao, Xing},
  title   = {Compact mid-infrared laser sensor for temperature and {CO} concentration measurement using heterodyne phase-sensitive dispersion spectroscopy},
  journal = {Measurement},
  volume  = {244},
  pages   = {116084},
  year    = {2025},
  doi     = {10.1016/j.measurement.2024.116084}
}

@article{Schulz2005,
  author  = {Schulz, Christof and Sick, Volker},
  title   = {Tracer-{LIF} diagnostics: quantitative measurement of fuel concentration, temperature and fuel/air ratio in practical combustion systems},
  journal = {Prog. Energy Combust. Sci.},
  volume  = {31},
  number  = {1},
  pages   = {75--121},
  year    = {2005},
  doi     = {10.1016/j.pecs.2004.08.002}
}

@article{Ehn2017,
  author  = {Ehn, Andreas and Zhu, Jiajian and Li, Xuesong and Kiefer, Johannes},
  title   = {Advanced Laser-Based Techniques for Gas-Phase Diagnostics in Combustion and Aerospace Engineering},
  journal = {Appl. Spectrosc.},
  volume  = {71},
  number  = {3},
  pages   = {341--366},
  year    = {2017},
  doi     = {10.1177/0003702817690161}
}

@article{Routh2013,
  author  = {Routh, Alexander F.},
  title   = {Drying of thin colloidal films},
  journal = {Rep. Prog. Phys.},
  volume  = {76},
  number  = {4},
  pages   = {046603},
  year    = {2013},
  doi     = {10.1088/0034-4885/76/4/046603}
}

@article{Law1982,
  author  = {Law, Chung K.},
  title   = {Recent advances in droplet vaporization and combustion},
  journal = {Prog. Energy Combust. Sci.},
  volume  = {8},
  number  = {3},
  pages   = {171--201},
  year    = {1982},
  doi     = {10.1016/0360-1285(82)90011-9}
}

@article{Plawsky2014,
  author  = {Plawsky, J. L. and Fedorov, A. G. and Garimella, S. V. and Ma, H. B. and Maroo, S. C. and Chen, L. and Nam, Y.},
  title   = {Nano- and Microstructures for Thin-Film Evaporation---A Review},
  journal = {Nanoscale Microscale Thermophys. Eng.},
  volume  = {18},
  number  = {3},
  pages   = {251--269},
  year    = {2014},
  doi     = {10.1080/15567265.2013.878419}
}

@article{Zhao2026,
  author  = {Zhao, Dongfang and Shi, Yumin and Wang, Shengkai},
  title   = {Spatiotemporally resolved measurements of {CO$_2$} distribution at the air--water interface using tunable diode laser spectroscopy},
  journal = {Phys. Rev. Fluids},
  volume  = {11},
  number  = {3},
  pages   = {034903},
  year    = {2026},
  doi     = {10.1103/j1df-csw4}
}

@article{Chen2023b,
  author  = {Chen, Xiaoliang and Zhu, Ning and Wakata, Yuki and Wang, Weitian and Song, Zihao and Chao, Xing},
  title   = {Full-field vapor concentration and temperature field measurement of an evaporating ethanol--water binary sessile drop by tomographic laser absorption spectroscopy},
  journal = {Phys. Fluids},
  volume  = {35},
  number  = {12},
  pages   = {123313},
  year    = {2023},
  doi     = {10.1063/5.0174342}
}

@article{Chen2023a,
  author  = {Chen, Xiaoliang and Zhu, Ning and Wang, Weitian and Wang, Zhenhai and Wakata, Yuki and Chao, Xing},
  title   = {Vapor concentration and temperature field measurement of an evaporating sessile drop by tomographic laser absorption spectroscopy},
  journal = {Phys. Fluids},
  volume  = {35},
  number  = {1},
  pages   = {012111},
  year    = {2023},
  doi     = {10.1063/5.0132210}
}

@article{Mi2026Leidenfrost,
  author  = {Mi, Yao and Chen, Xiaoliang and Wang, Weitian and Wakata, Yuki and Hu, Zhen and Chao, Xing},
  title   = {In situ sensing of temperature and concentration field in {Leidenfrost} vapor layers using tomographic laser absorption spectroscopy},
  journal = {Sens. Actuators B Chem.},
  volume  = {451},
  pages   = {139350},
  year    = {2026},
  doi     = {10.1016/j.snb.2025.139350}
}

@article{deegan1997capillary,
  title={Capillary flow as the cause of ring stains from dried liquid drops},
  author={Deegan, Robert D and Bakajin, Olgica and Dupont, Todd F and Huber, Greb and Nagel, Sidney R and Witten, Thomas A},
  journal={Nature},
  volume={389},
  number={6653},
  pages={827--829},
  year={1997},
  doi={10.1038/39827},
  publisher={Nature Publishing Group UK London}
}

@article{lohse2022fundamental,
  title={Fundamental fluid dynamics challenges in inkjet printing},
  author={Lohse, Detlef},
  journal={Annu. Rev. Fluid Mech.},
  volume={54},
  number={1},
  pages={349--382},
  year={2022},
  doi={10.1146/annurev-fluid-022321-114001},
  publisher={Annual Reviews}
}

@article{Dehaeck2014,
  author  = {Dehaeck, Sam and Rednikov, Alexey and Colinet, Pierre},
  title   = {Vapor-Based Interferometric Measurement of Local Evaporation Rate and Interfacial Temperature of Evaporating Droplets},
  journal = {Langmuir},
  year    = {2014},
  volume  = {30},
  number  = {8},
  pages   = {2002--2008},
  doi     = {10.1021/la404999z}
}

@misc{NISTPentane,
  author       = {{National Institute of Standards and Technology}},
  title        = {{NIST} {Chemistry WebBook}, {SRD} 69: Pentane},
  howpublished = {\url{https://webbook.nist.gov/cgi/cbook.cgi?ID=C109660}},
  year         = {2026},
  note         = {Accessed 8 September 2026}
}

@misc{NISTHexane,
  author       = {{National Institute of Standards and Technology}},
  title        = {{NIST} {Chemistry WebBook}, {SRD} 69: n-Hexane},
  howpublished = {\url{https://webbook.nist.gov/cgi/cbook.cgi?ID=C110543}},
  year         = {2026},
  note         = {Accessed 29 October 2026}
}

@misc{NISTHeptane,
  author       = {{National Institute of Standards and Technology}},
  title        = {{NIST} {Chemistry WebBook}, {SRD} 69: n-Heptane},
  howpublished = {\url{https://webbook.nist.gov/cgi/cbook.cgi?ID=C142825}},
  year         = {2026},
  note         = {Accessed 29 October 2026}
}

@article{Fuller1966,
  author  = {Fuller, Edward N. and Schettler, Paul D. and Giddings, J. Calvin},
  title   = {A New Method for Prediction of Binary Gas-Phase Diffusion Coefficients},
  journal = {Ind. Eng. Chem.},
  year    = {1966},
  volume  = {58},
  number  = {5},
  pages   = {18--27},
  doi     = {10.1021/ie50677a007}
}

@article{Mitrovic2012,
  author  = {Mitrovic, Jovan},
  title   = {Josef Stefan and his evaporation--diffusion tube---the Stefan diffusion problem},
  journal = {Chem. Eng. Sci.},
  volume  = {75},
  pages   = {279--281},
  year    = {2012},
  doi     = {10.1016/j.ces.2012.03.034}
}

@article{Vlasov2020,
  author  = {Vlasov, Valeriy A.},
  title   = {Diffusion-kinetic model of liquid evaporation from a Stefan tube: A solution to the Stefan diffusion problem},
  journal = {Int. J. Heat Mass Transf.},
  volume  = {163},
  pages   = {120379},
  year    = {2020},
  doi     = {10.1016/j.ijheatmasstransfer.2020.120379}
}

@article{Goldenstein2017IR,
  author  = {Goldenstein, Christopher S. and Spearrin, R. Mitchell and Jeffries, Jay B. and Hanson, Ronald K.},
  title   = {Infrared Laser-Absorption Sensing for Combustion Gases},
  journal = {Prog. Energy Combust. Sci.},
  volume  = {60},
  pages   = {132--176},
  year    = {2017},
  doi     = {10.1016/j.pecs.2016.12.002}
}

@article{Wei2018LAI,
  author  = {Wei, Chuyu and Pineda, Daniel I. and Goldenstein, Christopher S. and Spearrin, R. Mitchell},
  title   = {Tomographic Laser Absorption Imaging of Combustion Species and Temperature in the Mid-Wave Infrared},
  journal = {Opt. Express},
  volume  = {26},
  number  = {16},
  pages   = {20944--20955},
  year    = {2018},
  doi     = {10.1364/OE.26.020944}
}

@article{Schwarm2019LAI,
  author  = {Schwarm, Kevin K. and Wei, Chuyu and Pineda, Daniel I. and Spearrin, R. Mitchell},
  title   = {Time-Resolved Laser Absorption Imaging of Ethane at 2~kHz in Unsteady Partially Premixed Flames},
  journal = {Appl. Opt.},
  volume  = {58},
  number  = {21},
  pages   = {5656--5662},
  year    = {2019},
  doi     = {10.1364/AO.58.005656}
}

@article{Wei2024LAI,
  author  = {Wei, Chuyu and Spearrin, R. Mitchell},
  title   = {Four-Dimensional Laser Absorption Cinematography of Species and Temperature Dynamics at 2~kHz in Reacting Flows},
  journal = {Opt. Lett.},
  volume  = {49},
  number  = {1},
  pages   = {141--144},
  year    = {2024},
  doi     = {10.1364/OL.512240}
}

@article{Thorin2023,
  author  = {Thorin, Emil and Paiva, Eduardo M. and Schmidt, Florian M.},
  title   = {Quantitative Tomographic Laser Absorption Imaging of Atomic Potassium during Combustion of Potassium Chloride Salt and Biomass},
  journal = {Anal. Chem.},
  volume  = {95},
  number  = {2},
  pages   = {1140--1148},
  year    = {2023},
  doi     = {10.1021/acs.analchem.2c03890}
}

@article{Thorin2025,
  author  = {Thorin, Emil and Schmidt, Florian M.},
  title   = {Wide-Field Imaging of Gas-Phase {K} Species in Biomass Combustion Using Photofragmentation Laser Absorption Imaging},
  journal = {Fuel},
  volume  = {381},
  pages   = {133429},
  year    = {2025},
  doi     = {10.1016/j.fuel.2024.133429}
}

@article{Wang2026AlO,
  author  = {Wang, Weitian and Wu, Zhiyong and Song, Zihao and Chao, Xing and Ald{\'e}n, Marcus and Li, Zhongshan},
  title   = {{AlO} Evolution and Dynamics around Aluminum Droplets Burning in Water-Vapor-Rich Environments},
  journal = {Proc. Combust. Inst.},
  volume  = {42},
  pages   = {106171},
  year    = {2026},
  doi     = {10.1016/j.proci.2026.106171}
}

@article{Song2026Microjets,
  author  = {Song, Zihao and Wakata, Yuki and Zhao, Chaoben and Zhou, Quan and Sun, Chao and Chao, Xing},
  title   = {Optical Diagnostics of Flow Velocity in Pressure-Matched Near-Vacuum Gaseous Micro Jets},
  journal = {Sci. China Phys. Mech. Astron.},
  volume  = {69},
  number  = {5},
  pages   = {254511},
  year    = {2026},
  doi     = {10.1007/s11433-025-2880-3}
}

@article{Song2026Tin,
  author  = {Song, Zihao and Xu, Qi and Sun, Chao and Chao, Xing},
  title   = {Thermodynamic Scaling Laws and Ionization Kinetics of Laser-Produced Tin Plasmas at Various Pressures and Irradiating Laser Energies},
  journal = {Sci. China Phys. Mech. Astron.},
  volume  = {69},
  number  = {11},
  pages   = {114711},
  year    = {2026},
  doi     = {10.1007/s11433-026-3073-9}
}

@article{Adrian1991,
  author  = {Adrian, Ronald J.},
  title   = {Particle-Imaging Techniques for Experimental Fluid Mechanics},
  journal = {Annu. Rev. Fluid Mech.},
  volume  = {23},
  pages   = {261--304},
  year    = {1991},
  doi     = {10.1146/annurev.fl.23.010191.001401}
}

@article{Danehy2022FLEET,
  author  = {Danehy, Paul M. and Burns, Ross A. and Reese, Daniel T. and Retter, Jonathan E. and Kearney, Sean P.},
  title   = {{FLEET} Velocimetry for Aerodynamics},
  journal = {Annu. Rev. Fluid Mech.},
  volume  = {54},
  pages   = {525--553},
  year    = {2022},
  doi     = {10.1146/annurev-fluid-032321-025544}
}

@article{Borggren2017TLAF,
  author  = {Borggren, Jesper and Weng, Wubin and Hosseinnia, Ali and Bengtsson, Per-Erik and Ald{\'e}n, Marcus and Li, Zhongshan},
  title   = {Diode Laser-Based Thermometry Using Two-Line Atomic Fluorescence of Indium and Gallium},
  journal = {Appl. Phys. B},
  volume  = {123},
  pages   = {278},
  year    = {2017},
  doi     = {10.1007/s00340-017-6855-z}
}

@article{Wakata2023Leidenfrost,
  author  = {Wakata, Yuki and Zhu, Ning and Chen, Xiaoliang and Lyu, Sijia and Lohse, Detlef and Chao, Xing and Sun, Chao},
  title   = {How roughness and thermal properties of a solid substrate determine the {Leidenfrost} temperature: Experiments and a model},
  journal = {Phys. Rev. Fluids},
  volume  = {8},
  pages   = {L061601},
  year    = {2023},
  doi     = {10.1103/PhysRevFluids.8.L061601}
}

@article{Zeng2023,
  author  = {Zeng, Hao and Wakata, Yuki and Chao, Xing and Li, Mingbo and Sun, Chao},
  title   = {On evaporation dynamics of an acoustically levitated multicomponent droplet: Evaporation-triggered phase transition and freezing},
  journal = {J. Colloid Interface Sci.},
  volume  = {648},
  pages   = {736--744},
  year    = {2023},
  doi     = {10.1016/j.jcis.2023.06.012}
}

@article{Wakata2024,
  author  = {Wakata, Yuki and Chao, Xing and Sun, Chao and Diddens, Christian},
  title   = {Evaporation of acoustically levitated bicomponent droplets: Mass and heat transfer characteristics},
  journal = {J. Fluid Mech.},
  volume  = {984},
  pages   = {A17},
  year    = {2024},
  doi     = {10.1017/jfm.2024.204}
}

@article{Wakata2025,
  author  = {Wakata, Yuki and Wang, Feng and Sun, Chao and Lohse, Detlef},
  title   = {Thermal-solutal-induced bistability of evaporating multicomponent liquid thin films},
  journal = {Proc. Natl. Acad. Sci. U.S.A.},
  volume  = {122},
  number  = {6},
  pages   = {e2418487122},
  year    = {2025},
  doi     = {10.1073/pnas.2418487122}
}

@article{Wakata2026EIPS,
  author  = {Wakata, Yuki and Li, Mingbo and Sun, Chao},
  title   = {Evaporation-induced phase separation: Coupled mechanisms and applications},
  journal = {Curr. Opin. Colloid Interface Sci.},
  volume  = {81},
  pages   = {101987},
  year    = {2026},
  doi     = {10.1016/j.cocis.2025.101987}
}

@article{Li2025Darcy,
  author  = {Li, Junyi and Yang, Yantao and Sun, Chao},
  title   = {From {Darcy} convection to free-fluid convection: Pore-scale study of density-driven currents in porous media},
  journal = {J. Fluid Mech.},
  volume  = {1009},
  pages   = {A10},
  year    = {2025},
  doi     = {10.1017/jfm.2025.213}
}

@book{Bird2002,
  author    = {Bird, R. Byron and Stewart, Warren E. and Lightfoot, Edwin N.},
  title     = {Transport Phenomena},
  edition   = {2},
  publisher = {Wiley},
  address   = {New York},
  year      = {2002}
}

\end{document}